\documentclass[aps,prb,twocolumn,letterpaper,superscriptaddress,showpacs]{revtex4}
\usepackage{CJK}
\usepackage{keyval}%
\usepackage{graphicx}
\usepackage{dcolumn}
\usepackage{bm}
\usepackage{color}
\usepackage{amsmath}

\setkeys{Gin}{width=0.8\columnwidth,clip=true}
\newcommand{\ignore}[1]{}

\begin{document}
\begin{CJK*}{UTF8}{bsmi}
\title{Competing magnetism in $\pi$ electrons in graphene with a single carbon vacancy}
\author{Chi-Cheng Lee (%
李啟正
)}
\affiliation{School of Materials Science, Japan Advanced Institute of Science and Technology (JAIST),
1-1 Asahidai, Nomi, Ishikawa 923-1292, Japan}%
\author{Yukiko Yamada-Takamura
}
\affiliation{School of Materials Science, Japan Advanced Institute of Science and Technology (JAIST),
1-1 Asahidai, Nomi, Ishikawa 923-1292, Japan}%
\author{Taisuke Ozaki
}
\affiliation{School of Materials Science, Japan Advanced Institute of Science and Technology (JAIST),
1-1 Asahidai, Nomi, Ishikawa 923-1292, Japan}%
\affiliation{Research Center for Simulation Science, Japan Advanced Institute of Science and Technology (JAIST), 
1-1 Asahidai, Nomi, Ishikawa 923-1292, Japan}

\date{\today}

\begin{abstract}
One intriguing finding in graphene is the vacancy-induced magnetism that highlights the interesting interaction between local magnetic moments and conduction electrons. Within density functional theory, the current understanding of the ground state is that a Stoner instability gives rise to ferromagnetism of $\pi$ electrons aligned with the localized moment of a $\sigma$ dangling bond and the induced $\pi$ magnetic moments vanish at low vacancy concentrations. However, the observed Kondo effect suggests that $\pi$ electrons around the vacancy should antiferromagnetically couple to the local moment and carry non-vanishing moments. Here we propose that a phase possessing both significant out-of-plane displacements and $\pi$ bands with antiferromagnetic coupling to the localized $\sigma$ moment is the ground state. With the features we provide, it is possible for spin-resolved STM, STS, and ARPES measurements to verify the newly proposed phase. 
\end{abstract}

\pacs{61.48.Gh, 61.72.jd, 73.22.Pr, 75.75.-c}

\maketitle
\end{CJK*}

Tunable magnetism in two-dimensional materials is of great interest and is promising for technologies such as spintronics devices due to the minimized one-atom-thick functioning scale\cite{Geim,Yazyev}. Graphene, an exceptional representative of two-dimensional materials, can develop single carbon vacancies via irradiation experiments~\cite{Mizes,Ruffleux,Butz,Iijima,Urita}. With single vacancies created this way, the originally non-magnetic graphene has been demonstrated to possess local magnetic moments, without foreign elements\cite{Ugeda}. Inspired by this intriguing property and the potential applications in defect engineering, extensive studies have been performed on this topic\cite{Briddon,Nieminen,Helm,Zgierski,Ugeda2,Jianhao,Nanda,Palacios,Nair,Bin,Vojta,Casartelli,Freitas}. The nature of magnetism in the vacancy-induced magnetic moment was suggested to be spin-half by recent experiments and no long-range magnetic ordering can be identified\cite{Nair}. The antiferromagnetic-type Kondo effect has also been experimentally observed\cite{Jianhao}. However, the existence of a phase supporting the Kondo effect, which is expected for a magnetic impurity in a metallic system\cite{Anderson,Kondo}, has been lacking in the discussion of graphene with single vacancies within density functional theory (DFT). Although the out-of-plane displacement of the carbon atom carrying the dangling bond, which is needed to switch on the Kondo effect, has been investigated\cite{Vojta,Casartelli,Freitas}, the coupling between the conduction electrons and the localized magnetic moment of a $\sigma$ dangling bond and the structural modification supported by DFT have not been emphasized and still remain obscure. In fact, a phase with $\pi$ electrons antiferromagnetically coupled to the localized $\sigma$ moment has never been recognized as the ground state, in contrast to the preferred experimental interpretations\cite{Jianhao}. 

Studying the electronic structure of $\pi$ electrons in graphene is very challenging in DFT calculations even without explicit consideration of the strong correlation induced by the creation of vacancies, as in Kondo physics. Not only are the delicate $k$-point sampling and smearing width required to accurately describe the density of states at the Fermi energy but also the defect concentration could non-negligibly alter the calculated magnetic moment\cite{Helm,Palacios,Bin}. The scenario of the ground state best supported by DFT is a phase following the mechanism of Stoner instability\cite{Helm}, where the density of states of the quasi-localized $\pi$ electrons of one sublattice at the Fermi energy are spin polarized and the developed magnetic moment aligns with the localized moment of the $\sigma$ dangling bond favoring Hund's coupling. The $\pi$ electrons on the other sublattice result in smaller opposite magnetic moments via an additional exchange energy gain forming a ferrimagnetic pattern\cite{Helm}. Interestingly, the $\pi$ magnetic moments are predicted to vanish at any experimentally relevant vacancy concentration\cite{Palacios}. With various reported values of both magnetic moments and out-of-plane displacements, it is unclear if a new phase that can compete with the currently suggested ground state and can demonstrate a different magnetism on $\pi$ electrons has been overlooked. Obviously, it is interesting and timely to provide a more comprehensive picture of the possible phases in graphene with single vacancies for optimization of future defect engineering in graphene.

In this Letter, we introduce three competing phases in graphene with single vacancies by first-principles calculations. The three phases demonstrate different magnetism in $\pi$ electrons around the vacancy. Based on the magnetism of $\pi$ electrons with respect to that of the $\sigma$ dangling bond, the three distinct phases can be classified as ferromagnetic-dominant (F), antiferromagnetic-dominant (AF), and quenched antiferromagnetic (Q) phases. The F phase has been commonly calculated to be the ground state with a magnetic moment larger than one Bohr magneton per vacancy. However, our calculations reveal that the total energies of those phases are almost degenerate. Therefore, we argue that the total energy alone is not conclusive in determining the ground state but needs to be considered with the physical properties of the phases. Along this line, we submit that the AF phase is the ground state. The structure of the AF phase has not been recognized before. Importantly, the F and AF phases possess different features in spin density, density of states, and electronic band structures. If these qualitative features indicated by DFT can be verified by experiment, it would help settle the controversial issue of the ground state of graphene with single carbon vacancies at low concentrations. In the AF phase, we will introduce a large in-plane area showing significant out-of-plane displacements that would also constitute a stringent requirement for future theoretical calculations.

To study graphene with a single vacancy in a reasonably large supercell accompanied by a dense $k$-point sampling, we adopt a $10\times10\times1$ supercell with an $8\times8\times1$ $k$-point mesh and a $16\times16\times1$ supercell with a $5\times5\times1$ $k$-point mesh. The electronic temperature for smearing is set to 250 K. The distance between two layers is set to 20~\AA. DFT calculation with a generalized gradient approximation\cite{Kohn,Perdew} was performed using the OpenMX code based on norm-conserving pseudopotentials generated with multireference energies\cite{MBK} and optimized pseudoatomic basis functions\cite{Ozaki,openmx}. The experimental in-plane lattice constant in the calculation ($a$ = 2.4612~\AA) was chosen to represent a natural value in the low-concentration regime. For each carbon atom, two, two, and one optimized radial functions were allocated for the $s$-, $p$-, and $d$-orbitals, respectively\cite{supp}. A cut-off radius of 7 Bohr was chosen for the basis functions. The residual force on each atom was relaxed to be less than $6\times10^{-5}$ Hartree/Bohr.

Another important parameter is the number of real-space grids used for numerical integrations and for solution of the Poisson equation. The dense grids are expected to accurately describe the spin density and the corresponding energy gain via the out-of-plane displacements. A regular mesh of 390 Ry was adopted along the in-plane lattice vectors and 1727 Ry was chosen for the out-of-plane axis. For the completeness of basis functions around the vacancy, we allocated one set of carbon basis functions but without any contribution of the pseudopotential at the vacancy as a ghost atom in the $10\times10\times1$ supercell. We found the ghost atom can obtain a more complete antiferromagnetic screening of the localized moment of the $\sigma$ dangling bond in both the Q and AF phases to further reduce the magnetic moments, as listed in Table~\ref{tab:energy}. Since the energy sequence of the three phases is unchanged and the AF phase gains more energy than the F phase with the ghost atom, the ghost atom was not included in the $16\times16\times1$ supercell calculation for faster relaxation of atomic positions. 

\begin{table}[tbp]
\caption{\label{tab:energy}
Magnetic moments (M) and total energies (E) per vacancy of $10\times10\times1$ (199 atoms) and $16\times16\times1$ (511 atoms) supercell calculations are provided in $\mu_B$ and meV, respectively. The results including the ghost atoms (G) are also shown. The total energy of the AF phase is shifted to zero for each row.
}
\begin{ruledtabular}
\begin{tabular}{cccc}
                  &  F phase  &  Q phase  & AF phase  \\
\hline
M ($10\times10\times1$)  &  1.2969   &  1.0205   & 0.8824    \\
M with G ($10\times10\times1$)  &  1.2979   &  0.9741   & 0.8219    \\
M ($16\times16\times1$)  &  1.3194   &  0.9428   & 0.7797    \\
\hline
E ($10\times10\times1$)  &  17.501   &  12.924   & 0.0000     \\
E with G ($10\times10\times1$)  &  19.527   &  15.190   & 0.0000     \\
E ($16\times16\times1$)  &  15.103   &  11.019   & 0.0000    \\
\end{tabular}
\end{ruledtabular}
\end{table}

The magnetic moments and the total energies per vacancy of the three phases are given in Table~\ref{tab:energy}. Surprisingly, the F phase is always found to possess the highest energy, in contrast to current understanding. However, it should be noted that the energy differences among the three phases are quite small and approach the numerical noise. Any further improvement on the computational parameters or low temperatures could provide a comparable fluctuation in the total energy\cite{supp}. Additionally, it is unclear if the energy gain via the Kondo effect has been estimated correctly by the generalized gradient approximation within DFT. Up to this point, we can only conclude that the AF phase can compete strongly with the F phase in gaining total energy.

\begin{figure}[tbp]
\includegraphics[width=0.98\columnwidth,clip=true,angle=0]{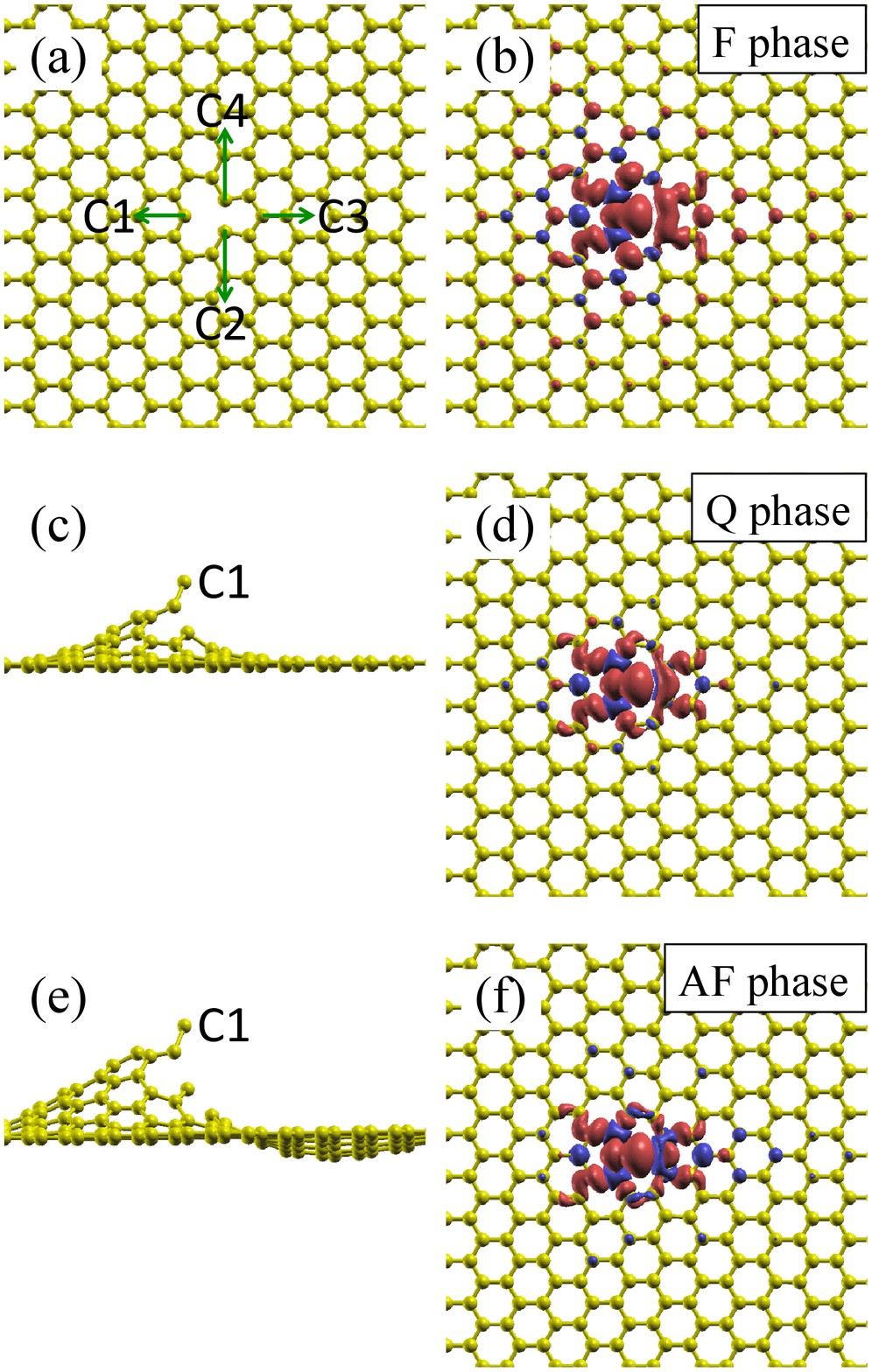}
\caption{\label{fig:fig1}
(a) In-plane structure of graphene with a single carbon vacancy. The lengths between C2 and C4 atoms are 1.86, 1.84, and 1.84~\AA~ in the F, Q, and AF phases, respectively. (b) Electron spin density of the F phase (Q phase in (d) and AF phase in (f)) is shown. The majority (positive) and minority (negative) spin density at 0.0003 e/Bohr$^3$ are colored in red and blue, respectively. The side views of the out-of-plane structures are presented for the (c) Q and (e) AF phases. The height of C1 atom measured from the C3 atom is 0.46~\AA~(0.68~\AA) in the Q (AF) phase. For a better visualization, the out-of-plane displacements in (c) and (e) are multiplied by ten. 
}
\end{figure}

The in-plane structure of all three phases shows the same Jahn-Teller distortion that reconstructs the bonding between the C2 and C4 atoms defined in Fig.~\ref{fig:fig1} (a). It would be difficult to experimentally distinguish one phase from another, however, distinct out-of-plane structures exist among the three phases. The F phase is planar, while the Q phase has a protruding C1 atom. The AF phase exhibits an interesting structure with a protruding C1 atom that forms a hill shape with its neighbors in addition to a basin shape beside the hill. This peculiar structure involves out-of-plane displacements over a large area. In fact, a noticeable wavy shape can still be found at the boundary of the $10\times10\times1$ supercell, indicating that a larger supercell is needed to avoid strong vacancy-vacancy interactions in the AF phase. In the $16\times16\times1$ supercell, the wavy boundary becomes unapparent. The structures calculated in the $16\times16\times1$ supercell are shown in Fig.~\ref{fig:fig1}.

Besides the geometrical structure, the F and AF phases show a striking difference in spin density that can be measured by a local probe, such as in a spin-resolved scanning tunneling microscope (STM) experiment. The F phase follows the known scenario that the $\pi$ electron spin density is positive in one sublattice and negative in the other, as shown in Fig.~\ref{fig:fig1} (b)\cite{Helm}. However, the spin density in the AF phase does not follow this pattern. A prominent difference from the F phase is the negative spin density on the C3 atom, as shown in Fig.~\ref{fig:fig1} (f). In the Q phase, the same negative spin density on the C3 atom occurs but the overall antiferromagnetism of $\pi$ electrons are quenched, as shown in Fig.~\ref{fig:fig1} (d). Obviously, the newly induced antiferromagnetism originates from the out-of-plane displacements, mainly from hopping between the $sp^2$ orbital of the C1 atom to the $p_z$ orbitals of the C2 and C4 atoms. This in turn changes the antiferromagnetism on the neighboring atoms, which competes with the existing magnetism. The significant out-of-plane displacements in the AF phase maximize the total energy gain involving both the ion positions and the local spin density of electrons in comparison to the less prominent displacements in the Q phase. Note that the displacements are not introduced by any external strain since the lattice constants are fixed at experimental values.

From the calculated magnetic moments and the spin density among the three phases, it is clear that the ferrimagnetic $\pi$ electrons tend to polarize the majority of the spin density to align with the localized moment of the $\sigma$ dangling bond ($m$ = 1 $\mu_B$) in gaining energy via Hund's coupling. Therefore, the total magnetic moment per vacancy is larger than 1 $\mu_B$. However, this ferrimagnetism of $\pi$ electrons in the AF phase competes with the magnetism coming from the new coupling between the localized $\sigma$ moment and the $\pi$ electrons via the Kondo-like effect after forming the hill-basin structure. As a result, the negative spin density of $\pi$ electrons in the AF phase activates a magnetic screening of the localized $\sigma$ moment and the magnetic moment per vacancy is found to be less than 1 $\mu_B$. The intermediate Q phase also tends to screen the localized $\sigma$ moment via the negative $\pi$ spin density, but antiferromagnetism via the Kondo-like effect is quenched due to suppressed out-of-plane displacements.

When the cell size is increased from $10\times10\times1$ to $16\times16\times1$, the moment per vacancy in the F phase does not decrease, rather, the moment seems to converge. This non-vanishing magnetism in $\pi$ electrons is consistent with the finding in Ref.~\onlinecite{Bin}. Since the largest size we have explored is $16\times16\times1$, we cannot rule out the possibility of vanishing $\pi$ magnetism at larger cell sizes. We presume that the non-vanishing $\pi$ magnetism is biased by positive spin density on the $\sigma$ dangling bond via Hund's coupling. Therefore, vanishing $\pi$ magnetism at a much lower vacancy concentration requires another competitor to modify the $\pi$ spin density. For example, non-zero hopping between the $sp^2$ orbital of the C1 atom and the $p_z$ orbitals of the C2 and C4 atoms could diminish the positive $\pi$ spin density on the atoms surrounding the vacancy. On the other hand, the magnetic moment per vacancy in the AF phase was shown to decrease with increased supercell size, with a trend toward a singlet state. The enhanced antiferromagnetic screening also suppresses long-range magnetic ordering. The smaller magnetic moment and suppressed long-range order in the AF phase are in good agreement with experimental findings\cite{Nair}.

\begin{figure}[ht]
\includegraphics[width=0.96\columnwidth,clip=true,angle=0]{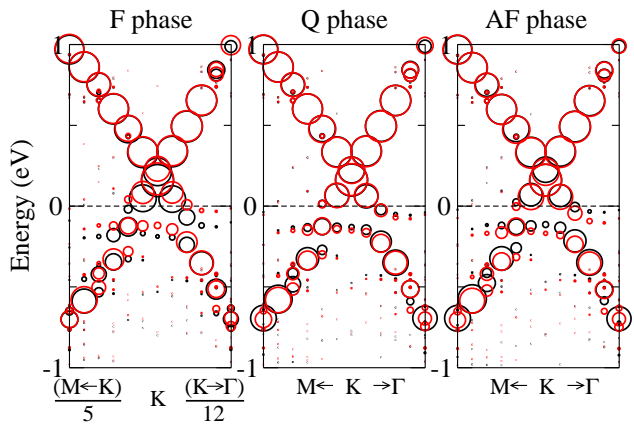}
\caption{\label{fig:fig2}
(color online). The electronic band structures of the F, Q, and AF phases in the representation of the Brillouin zone corresponding to the primitive unit cell without any vacancy. The spectral weight is represented by the diameter of a circle. The spin majority is colored in black and the spin minority is colored in red (gray). 
}
\end{figure}

The measured Kondo effect in resistivity suggests that the $\pi$ electrons should antiferromagnetically couple to the localized moment of the $\sigma$ dangling bond below the Kondo temperature\cite{Jianhao}. This suggests that the electronic band structure of the ground state in DFT should demonstrate split $\pi$ bands around the Fermi energy that are antiferromagnetically coupled to the majority spin. Electronic energy gain can be obtained by lowering the bonding-type $\sigma$ band coupled to the anti-bonding-type $\pi$ band in the spin majority channel. This can be found in the subsequent discussion of local density of states. In a global probe such as a resistivity measurement, the vacancy state itself could be negligible in the low-concentration limit because of the possibly negligible quasi-particle lifetime of the vacancy state in comparison with conduction electrons possessing longer mean free paths. To illustrate the electronic band structures of the three phases with the proper quasi-particle lifetime, we calculate the band structures by representing the spectral weight in the Brillouin zone of two carbon atoms\cite{Wei,Lee}, which is the primitive unit cell without any vacancy. The results for a $16\times16\times1$ supercell are shown in Fig.~\ref{fig:fig2}.

\begin{figure}[hb]
\includegraphics[width=0.96\columnwidth,clip=true,angle=0]{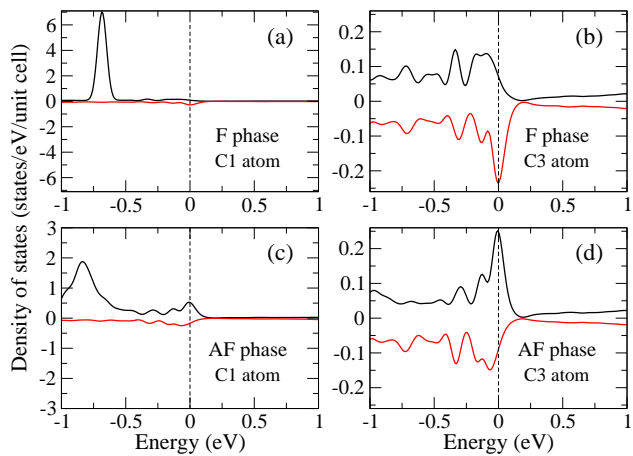}
\caption{\label{fig:fig3}
(color online). Local density of states of the (a) C1 and (b) C3 atoms in the F phase compared to the (c) C1 and (d) C3 atoms in the AF phase with a Gaussian broadening of 0.05 eV. The contributions of the C1 and C3 atoms in this energy range are mainly from the $sp^2$ and $p_z$ orbitals, respectively. The spin-majority density of states is presented by the positive value while the negative value is chosen for the spin-minority density of states. 
}
\end{figure}

Fig.~\ref{fig:fig2} clearly shows that the $\pi$ bands around the Fermi energy are ferromagnetically and antiferromagnetically spin-polarized to the spin-majority moment of the $\sigma$ dangling bond [cf. Fig.~\ref{fig:fig3} (a) and (c)] in the F and AF phases, respectively. Even without the significant out-of-plane displacements, the Q phase also follows the trend found in the AF phase. Although our studies suffer from the periodic boundary condition, the distinct qualitative behaviors should enable verification by spin- and angle-resolved photoemission spectroscopy (ARPES) measurements around the $K$ point. Since the antiferromagnetic behavior shown in $\pi$ electrons is consistent with the resistivity measurements~\cite{Jianhao}, the AF phase is suggested as the ground state. 

We want to highlight that one can observe spin splitting involving a large amount of $\pi$ electrons with longer quasi-particle lifetimes, as shown in this representation. Therefore, band splitting is expected to be smaller to maintain a fixed vacancy-induced magnetic moment in integrating the difference between the spin-majority and spin-minority density of states as the supercell size increases. This is consistent with the behavior reported in Ref.~\onlinecite{Palacios}. Finally, we plot the density of states of C1 and C3 atoms in both F and AF phases in Fig.~\ref{fig:fig3}. The opposite spin contributions of C3 atoms at the zero energy (Fermi energy) also provide good fingerprints for a local experimental probe, like spin-resolved scanning tunneling spectroscopy (STS), to confirm the existence of the proposed AF phase. Note that the lowered energy of the majority spin states shown in Fig.~\ref{fig:fig3} (c) reflects the electronic energy gain of the C1 atom in the AF phase. 

In conclusion, a new phase possessing conduction $\pi$ electrons antiferromagnetically coupled to the localized magnetic moment of the $\sigma$ dangling bond has been introduced by first-principles calculations and proposed as the ground state of graphene with a single carbon vacancy. This phase has a significant out-of-plane structural arrangement that maximizes total energy gain involving complicated $sp^2-p_z$ hopping in a large in-plane area. Such a non-planar structure is difficult to find in pristine graphene. When hopping is switched on, a fierce competition is triggered between the Kondo-like magnetism and the existing ferrimagnetism of $\pi$ electrons found in the planar structure. The large-scale modification in real space also sets a stringent constrain for future quantitative studies by DFT and model calculations. The DFT calculations reveal that the spin density, density of states, and band structure of the new phase are distinct from those of the previously reported ground state. These features should allow the existence of the new phase to be experimentally verified by methods such as spin-resolved STM, STS, and ARPES measurements. We expect that our findings, along with future experimental verifications, will greatly advance the understanding of graphene with single vacancies to optimize applications in spin electronics, defect engineering, and other graphene-related applications.

\section*{Acknowledgements}
This work was supported by the Strategic Programs for Innovative Research (SPIRE), MEXT, the Computational Materials Science Initiative (CMSI), and Materials Design through Computics: Complex Correlation and Non-Equilibrium Dynamics, A Grant in Aid for Scientific Research on Innovative Areas, MEXT, Japan. The calculations were performed using the Cray XC30 machine at Japan Advanced Institute of Science and Technology (JAIST).

\bibliography{refs}
\end{document}